\documentclass[prd,twocolumn,showpacs,floatfix,amsmath,amssymb,floatfix]{revtex4}
\usepackage{graphicx,color,dcolumn,booktabs,bm}
\usepackage{longtable,lscape}
\usepackage{txfonts}
\usepackage{overpic}
\usepackage{amssymb}
\usepackage{indentfirst}

\begin{document}

\title{Production of charmed baryon $\Lambda_c(2940)^+$ at PANDA}
\author{Jun He$^{1,3}$}
\author{Zhen Ouyang$^{1,3}$}
\author{Xiang Liu$^{1,2}$\footnote{Corresponding author}}\email{xiangliu@lzu.edu.cn}
\affiliation{
$^1$Research Center for Hadron and CSR Physics,
Lanzhou University and Institute of Modern Physics of CAS, Lanzhou 730000, China\\
$^2$School of Physical Science and Technology, Lanzhou University, Lanzhou 730000,  China\\
$^3$Nuclear Theory Group, Institute of Modern Physics of CAS, Lanzhou 730000, China
}
\author{Xue-Qian Li}
\affiliation{School of Physics, Nankai University, Tianjin 300071, China}
\date{\today}

\begin{abstract}

In this work we evaluate the production rate of the charmed baryon
$\Lambda_c(2940)^+$ at PANDA. For possible assignments of
$\Lambda_c(2940)^+$: $J^P=1/2^\pm$, $3/2^\pm$ and $5/2^\pm$, the
total cross section of $p\bar{p}\to \bar{\Lambda}_c
\Lambda_c(2940)^+$ is estimated, which may exceed 1 nb. With the
designed luminosity ($2\times10^{-32}$cm$^{-2}$/s) of PANDA,  our
estimate indicates that ten thousand events per day if
$\Lambda_c(2940)^+$ is of $J^P=1/2^+$ or $10^8$ per day if it is
of $J^P=5/2^+$ can be expected. Those values actually set the
lower and upper limits of the $\Lambda_c(2940)^+$ production. In
addition, we present the Dalitz plot and carry out a rough
background analysis of the $\Lambda_c(2940)^+$ production in the
$p\bar{p}\to  D^0 p\bar{\Lambda}_c$ and $p\bar{p}\to
\Sigma_c^{0,++}\pi^{+,-}\bar{\Lambda}_c$ processes, which would
provide valuable information for accurate determination of the
$\Lambda_c(2940)^+$ identity.
\end{abstract}

 \maketitle
\maketitle

\section{Introduction}\label{sec1}

The charmed baryon $\Lambda_c(2940)^+$ with mass $m=2939.8\pm1.3(\mathrm{stat})\pm1.0(\mathrm{syst})$ MeV
and width $\Gamma=17.5\pm5.2(\mathrm{stat})\pm5.9(\mathrm{syst})$ MeV
was first observed in the $D^0p$ invariant mass spectrum by the BaBar
Collaboration \cite{Aubert:2006sp}. Later, $\Lambda_c(2940)^+$ was confirmed by the Belle
Collaboration in
the $\Sigma_c(2455)^{0,++}\pi^{+,-}$ channels \cite{Abe:2006rz}, where the obtained mass and width are
$m=2938.0\pm1.3^{+2.0}_{-4.0}$ MeV and
$\Gamma=13^{+8+27}_{-5-7}$ MeV respectively. Obviously the values achieved by the two collaborations are consistent with each other within the error tolerance \cite{Aubert:2006sp}.

Actually, comparing with the meson case, the structure of baryons
is more intriguing from both theoretical and experimental aspects.
Recently, along with the experimental progress at the BaBar, Belle
and BES, a great number of new states of mesons have been observed
and some of them are identified as exotic, i.e., these states
cannot be categorized into the regular $q\bar q^\prime$ structure.
It is natural to conjecture that the possibility also exists for
the baryons. However, this situation is much more complicated than
the meson case. By the regular structure, the baryon is composed of
three quarks, so the exotic configuration of baryons would be much
more difficult to be identified. On the other side, this study can
enrich our knowledge on the fundamental structure of hadrons;
namely, it will answer the long-standing question that the $SU(3)$
theory indeed allows existence of the non $q\bar q$ and $qqq$
configurations, and, if yes, where do we search for them? That is the
job of theorists of high energy physics.

Experimentally, some peculiar phenomena have been observed. Before we can  attribute them to new physics or new hadronic configuration,
a thorough study of whether they can be interpreted by the regular quark structure and the standard model (SM) must be carried out.

The observation of $\Lambda_c(2940)^+$ has stimulated theorists'
extensive interest in understanding its structure. Since the
observed charmed baryon $\Lambda_c(2940)^+$ is close to the
production threshold of $D^*p$, a conjecture that
$\Lambda_c(2940)^+$ may be a $D^*N$ molecular state, was naturally
proposed \cite{He:2006is}. The masses of $D^*N$ molecular states
were calculated in the potential model, and the results support
the statement that $\Lambda_c(2940)^+$ is an S-wave $D^{*0}p$
molecular state with spin parity $J^{P}=\frac{1}{2}^-$ or
$J^{P}=\frac{1}{2}^+$ \cite{He:2006is}. Recently, the authors of
Ref. \cite{He:2010zq} systematically studied the interaction
between $D^*$ and the nucleon, and concluded that the $D^*N$ systems
may behave as  $J^P=1/2^\pm,\,3/2^\pm$ baryon states. With the
$J^{P}=\frac{1}{2}^-$ and $J^{P}=\frac{1}{2}^+$ assignments, the
strong decays of $\Lambda_c(2940)^+$ have been investigated by the
authors of Ref. \cite{Dong:2009tg}, but their result determines
that the assignment of $\Lambda_c(2940)^+$ as a $D^*N$ molecular
state with $J^{P}=\frac{1}{2}^-$ should be excluded. Later, the
radiative and strong three-body decays of $\Lambda_c(2940)^+$ were
explored in Refs. \cite{Dong:2010xv,Dong:2011ys}, where
$\Lambda_c(2940)^+$ was assigned as a $D^*N$ molecular state of
$J^{P}=\frac{1}{2}^+$.

Besides supposing $\Lambda_c(2940)^+$ to be a molecular state, the
alternative theoretical explanation that $\Lambda_{c}(2940)^+$ is
just a conventional charmed baryon has also been widely discussed.
The calculation in terms of the potential model shows that the
masses of the conventional charmed baryons of
$J^{P}=\frac{5}{2}^-$ and  $J^{P}=\frac{3}{2}^+$ are 2900 MeV and
2910 MeV, respectively \cite{Capstick:1986bm,Copley:1979wj}, which
are close to the mass of $\Lambda_c(2940)^+$. In Ref.
\cite{Ebert:2007nw}, the authors suggested that
$\Lambda_c(2940)^+$ is the first radial excitation of
$\Sigma_c(2520)$ of $J^P=\frac{3}{2}^+$ and possesses the quantum
number of $J^P=\frac{3}{2}^+$. In their calculations of the mass
spectrum the relativistic quark-diquark model was used. In
addition, $\Lambda_c(2940)^+$ as the first radial excitation of
the $\Sigma_c$ was also suggested via solving the  Faddeev
equations for three-body systems in the momentum space
\cite{Valcarce:2008dr}. In the heavy hadron chiral perturbation
theory, the ratio
$\Gamma(\Lambda_{c}(2940)^+\to\Sigma_c^*\pi)/\Gamma(\Lambda_{c}(2940)^+\to\Sigma_c\pi)$
was obtained if the spin-parity of $\Lambda_{c}(2940)^+$ is
$J^{P}=\frac{5}{2}^-$ or $J^{P}=\frac{3}{2}^+$
\cite{Cheng:2006dk}. These ratios will be applied to distinguish
different $J^P$ assignments of $\Lambda_{c}(2940)^+$
\cite{Cheng:2006dk}. In Ref. \cite{Chen:2007xf}, the authors
calculated the strong decays of newly observed charmed hadrons in
the $^3P_0$ model. Here, $\Lambda_{c}(2940)^+$ could only be a
D-wave charmed baryon $\check{\Lambda}_{c1}^0(\frac{1}{2}^+)$ or
$\check{\Lambda}_{c1}^0(\frac{3}{2}^+)$ while
$\Lambda_{c}(2940)^+$ as the first radial excitation of
$\Lambda_c(2286)^+$ is completely excluded since
$\Lambda_{c}(2940)^+\to D^0 p$ was observed by the BaBar
Collaboration \cite{Aubert:2006sp}. The result obtained in terms
of the chiral quark model indicates that $\Lambda_c(2940)^+$ could
be a D-wave charmed baryon
$\Lambda_c\,^2D_{\lambda\lambda}\frac{3}{2}^+$
\cite{Zhong:2007gp}.

\begin{table}[htb]
    \caption{The possible $J^P$ assignments to the
    $\Lambda_c(2940)^+$ in the literature \cite{He:2006is,Dong:2009tg,Dong:2010xv,Dong:2011ys,He:2010zq,Capstick:1986bm,Copley:1979wj,Cheng:2006dk,
Zhong:2007gp,Chen:2007xf,Ebert:2007nw,Valcarce:2008dr,Chen:2009tm}.
Here, we use "$\checkmark$" or "$\times$" to denote that the corresponding studies suggest or exclude that $J^P$ assignment for $\Lambda_c(2940)^+$.
Additionally, the upper and lower values in the bracket denote the decay
    widths (MeV) for its $D^0p$ and $\Sigma_c^{++}\pi^-$
    channels obtained in the literature corresponding to the quantum number assignments.\label{Tab:ref}}
\small
\renewcommand\tabcolsep{0.08cm}
\begin{center}
    \begin{tabular}{ ll | cccccc}  \toprule[1pt]
    &  &      $1/2^+$ & $1/2^-$ & $3/2^+$ &$3/2^-$& $5/2^+$ &$5/2^-$\\\midrule[1pt]
He {et al.}   &\cite{He:2006is}   &  & $\checkmark$         &        &$\checkmark$       &        & \\
Dong {et al.}&\cite{Dong:2009tg}   &(\small$^{0.20\pm0.09}_{0.95\pm0.37}$)   &$\times$         &        &       &        & \\
Dong {et al.}&\cite{Dong:2010xv,Dong:2011ys}   &$\checkmark$        &         &        &       &        & \\
He {et al.}  &\cite{He:2010zq}   &$\checkmark$   &         &        &$\checkmark$       &
& \\\midrule[1pt]
Capstick {et al.}&\cite{Capstick:1986bm,Copley:1979wj} &       &         &$\checkmark$        &       &$\checkmark$        &$\checkmark$\\
Cheng {et al.}  &\cite{Cheng:2006dk}   &        &         & $\checkmark$       &       &        &$\checkmark$ \\
Zhong {et al.}  &\cite{Zhong:2007gp}   &        &         &        &
&($^{1.08}_{1.06}$)        & \\
Chen {et al.}   &\cite{Chen:2007xf}   &($^{11}_{2.2}$)        &         &
($^{11}_{0.6}$)       &       &        & \\
Ebert {et al.}  &\cite{Ebert:2007nw}   &        &         & $\checkmark$       &       &        & \\
Valcarce {et al.}&\cite{Valcarce:2008dr}   &        &         & $\checkmark$       &       &        & \\
Chen {et al.}&\cite{Chen:2009tm}   &   &         &        &       &
&$\checkmark$\\
\bottomrule[1pt]
\end{tabular}
\end{center}
\end{table}

As summarized in Table ~\ref{Tab:ref}, a great deal of theoretical
ansatz for the structure $\Lambda_c(2940)^+$ was proposed, by which its spectrum was calculated, and
the results are quite model dependent. At present
the properties of $\Lambda_c(2940)^+$ are still unclear, the fact means
that more work is needed to determine its real structure, especially investigating from different angles.

The current information of $\Lambda_c(2940)^+$ is extracted from
the $e^+e^-$ collision \cite{Aubert:2006sp}. Thus, it is
interesting to investigate the $\Lambda_c(2940)^+$ production in
other processes. The PANDA experiment \cite{Lutz:2009ff} at the
Facility for Antiproton and Ion Research (FAIR) will be carried
out in the near future, which will definitely provide valuable
data for understanding of non-perturbative QCD. Study of the charmed
baryon is one of the main physics goals of PANDA since its  beam
momentum $p=5\sim15$ GeV just covers the production threshold of
charmed hadron. Encouraged by the prospect, in this work, we study
the $\Lambda_c(2940)^+$ production at PANDA. Some parallel
theoretical investigations of the production of the
charminium-like states $X(3872)$, $Z^+(4430)$ at
PANDA\cite{Chen:2008cg,Ke:2008kf} were also carried out.

This paper is organized as follows. After the Introduction, we
will present the effective Lagrangian and the corresponding
coupling constants used in this work. The formulation and the
numerical result of the $\Lambda_c(2940)^+$ productions at PANDA
will be given in Sec.~\ref{2940}. In Sec.~\ref{background},
considering the sequential decay $\Lambda_c(2940)^+\to D^0p$, we
make the Dalitz plot analysis on $p\bar{p}\to
\bar{\Lambda}_c\Lambda_c(2940)^+\to \bar{\Lambda}_c D^0 p$, where
$p\bar{p}\to \bar{\Lambda}_c\Lambda_c\to \bar{\Lambda}_c D^0 p$
forms the background. Finally the paper ends with our discussion
and conclusion.

\section{Effective Lagrangians and coupling constant}

\begin{figure}[htb]
\centering
\begin{tabular}{c}
\scalebox{0.85}{\includegraphics{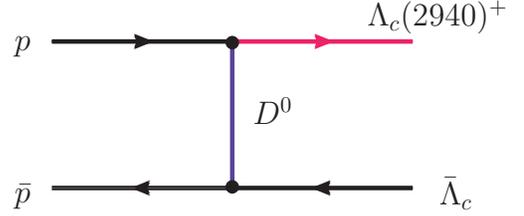}}
\end{tabular}
\caption{The diagram for the process
$p\bar{p}\to \bar{\Lambda}_c\Lambda_c(2940)^+$ by exchanging the $D^0$
meson.\label{diagram22}}
\end{figure}

Associated with a $\bar{\Lambda}_c$ production, $\Lambda_c(2940)^+$ could be produced in the proton and antiproton collision by exchanging
a $D^0$ meson, as shown in the Fig.~\ref{diagram22}. It is noted that direct $p\bar{p}$ annihilation into $\bar{\Lambda}_c\Lambda_c(2940)^+$
is negligible in comparison with the mechanism shown in Fig.~\ref{diagram22}, because
the annihilation channel is Okubo-Zweig-Iizuka (OZI) suppressed. Thus, in this work we do not consider its contribution at all.

For being at most model-independent, we apply the effective Lagrangian approach to study the $p\bar{p}\to \bar{\Lambda}_c\Lambda(2940)^+$ process.
In our calculation, we consider the production rates of $\Lambda_c(2940)^+$ whose $J^P$ assignments are priori assumed.
The following Lagrangians describe the interaction of $\Lambda_c(2940)^+$ with $D^0 p$ for different $J^P$ assignments to $\Lambda_c(2940)^+$
\cite{Oh:2007ej,Sibirtsev2000,Haglin1999,Dong:2009tg,Wu:2009md}:
\begin{eqnarray}
{\cal
L}_{\frac{1}{2}^+}&=&g_{\frac{1}{2}^+}~{\Lambda}_c(2940)^+~i\gamma_5~p~D^0,\label{L1}\\
{\cal
L}_{\frac{1}{2}^-}&=&g_{\frac{1}{2}^-}~{\Lambda}_c(2940)^+~p~D^0,\\
{\cal
L}_{\frac{3}{2}^+}&=&g_{\frac{3}{2}^+}~{\Lambda}_c^\mu(2940)^+~p~\partial_\mu
D^0,\\
{\cal
L}_{\frac{3}{2}^-}&=&g_{\frac{3}{2}^-}~{\Lambda}_c^\mu(1940)^+~i\gamma_5~p~\partial_\mu
D^0,\\
{\cal
L}_{\frac{5}{2}^+}&=&g_{\frac{5}{2}^+}~{\Lambda}_c^{\mu\nu}(2940)^+~i\gamma_5~
p~\partial_\mu\partial_\nu D^0,\\
{\cal
L}_{\frac{5}{2}^-}&=&g_{\frac{5}{2}^-}~{\Lambda}_c^{\mu\nu}(2940)^+~p~\partial_\mu\partial_\nu
D^0,\label{L6}
\end{eqnarray}
where we use the subscripts $\frac{1}{2}^\pm$, $\frac{3}{2}^\pm$ and $\frac{5}{2}^\pm$ to distinguish possible $J^P$ quantum numbers of $\Lambda_c(2940)^+$. The Lagrangian for the interaction of $\bar{\Lambda}_c$ and $\bar{D}^0\bar{p}$ can
be easily obtained by  replacing $\Lambda_c(2940)^+$ ($p,D^0$) in Eq. (\ref{L1}) with $\bar{\Lambda}_c$ ($\bar{p},\bar{D}^0$). In the above Lagrangians, the coupling constants $g_{_{J^P}}\equiv g_{\Lambda_c(2940)^+pD^0}$
can be obtained by fitting the measured partial width of the $\Lambda_c(2940)^+\to D^0p$ decay, i.e.,
\begin{eqnarray}
    {{\Gamma(\Lambda_c(2940)^+ \to pD^0)}\over{g_{_{J^P}}^2}}=\frac{m_N}{4(2J+1)\pi }
    \frac{2|\bm k|}{\sqrt{s}}B_{ {\cal S}}A^{J}
\end{eqnarray}
with $B_{\cal S}=\frac{E_N}{m_N} +{\cal S}$ and ${\cal
S}=P(-1)^{J+1/2}$, where $J$ is the spin of $\Lambda_c(2940)$, $E_N$ ($m_N$) denotes the energy (mass) of proton.
$A^J=\frac{N}{2J}|\bm k|^{2J-1}$ with $N=1, 2, 2$ corresponds to $J=1/2, 3/2, 5/2$, respectively. $\bm k$ is the three-momentum of the daughter mesons in the  center of mass frame of $p\bar p$.
From $BR(\Lambda_c(2940)^+\to D^0p)$, we extract the coupling constant $g_{_{J^P}}$. However, the BaBar and Belle experiments only measured the total width of $\Lambda_c(2940)^+$, and have
not given the partial decay width of $\Lambda_c(2940)^+\to D^0p$ so far.
Thus, to obtain $g_{_{J^P}}$, one needs to invoke theoretical calculations. In terms of different theoretical models to estimate, different groups have obtained different values of the decay width of $\Lambda_c(2940)^+\to D^0p$
which are listed in Table. \ref{Tab:ref}.
Since the cross section of $p\bar{p}\to \bar{\Lambda}_c\Lambda_c(2940)^+$ is proportional to $g_{_{J^P}}^2$, the line shape of the cross section of
$p\bar{p}\to \Lambda_c(2940)^+\bar{\Lambda}_c$ depends on the c.m. energy $\sqrt{s}$, but does not depend on the $g_{_{J^P}}$ value.
In this work, we choose a concrete $g_{_{J^P}}$ value to calculate the cross section of $p\bar{p}\to \bar{\Lambda}_c\Lambda_c(2940)^+$.
Concretely, we set the partial decay width  to be $\Gamma(\Lambda_c(2940)^+\to D^0 p)=1.5$ MeV and then determine the coupling constant $g_{_{J^P}}$ as
$g_{_{\frac{1}{2}^-}}=0.26$; $g_{_{\frac{1}{2}^+}}=1.25$; $g_{_{\frac{3}{2}^-}}=5.26$ GeV$^{-1}$; $g_{_{\frac{3}{2}^+}}=1.10$ GeV$^{-1}$;
$g_{_{\frac{5}{2}^-}}=4.23$ GeV$^{-2}$ and $g_{_{\frac{5}{2}^+}}=20.19$ GeV$^{-2}$.
By an approximate $SU(4)$ flavor symmetry, the coupling constant
$g_{\Lambda_c pD^0}$ is equal to $g_{\Lambda NK }=13.2$
\cite{Rijken:1998yy,Stoks:1999bz,Oh:2006hm,Liu:2010um}, which is
larger than $g_{\Lambda NK }=6.7\pm2.1$ estimated via the QCD sum
rules \cite{Navarra:1998vi,Bracco:1999xe}.

The propagators for a fermion of $J=1/2, 3/2, 5/2$ are written
as~\cite{Huang:2005js,Wu:2009md},
\begin{eqnarray}
    G^{n+\frac{1}{2}}(q)&=&P^{(n+\frac{1}{2})}G_R(q^2)\nonumber\\
    &=&P^{(n+\frac{1}{2})}
    \frac{2M_R}{q^2-M^2_R+iM_R\Gamma_R}
\end{eqnarray}
with
\begin{eqnarray}
    P^\frac{1}{2}(q)&=&\frac{\rlap\slash q+M_R}{2M_R},\\
    P^\frac{3}{2}_{\mu\nu}(q)
    &=&\frac{\rlap\slash q+M_R}{2M_R}
     \Bigg[-g_{\mu\nu}+\frac{1}{3}\gamma_\mu\gamma_\nu+\frac{1}{3M_R}(\gamma_\mu q_\nu-\gamma_\nu q_\mu)
    \nonumber\\
    &&+\frac{2}{3M_R^2}q_\mu q_\nu\Bigg],\\
    P^\frac{5}{2}_{\mu_1\mu_2\nu_1\nu_2}(q)
    &=&\frac{\rlap\slash q+M_R}{2M_R}\Bigg[\frac{1}{2}(\widetilde{g}_{\mu_1\nu_1}
    \widetilde{g}_{\mu_2\nu_2}+\widetilde{g}_{\mu_1\nu_2}
    \widetilde{g}_{\mu_2\nu_1})\nonumber\\
    &&-\frac{1}{5}\widetilde{g}_{\mu_1\mu_2}
    \widetilde{g}_{\nu_1\nu_2}-\frac{1}{10}
     (\widetilde{\gamma}_{\mu_1}\widetilde{\gamma}_{\nu_1}\widetilde{g}_{\mu_2\nu_2}
     +\widetilde{\gamma}_{\mu_1}\widetilde{\gamma}_{\nu_2}\widetilde{g}_{\mu_2\nu_1}\nonumber\\
     &&+\widetilde{\gamma}_{\mu_2}\widetilde{\gamma}_{\nu_1}\widetilde{g}_{\mu_1\nu_2}
     +\widetilde{\gamma}_{\mu_2}\widetilde{\gamma}_{\nu_2}\widetilde{g}_{\mu_1\nu_1})\Bigg],
\end{eqnarray}
where $\widetilde{\gamma}_\nu=\gamma_\nu-q_\nu\rlap\slash q/q^2$
and $\widetilde{g}_{\mu\nu}=g_{\mu\nu}-q_\mu q_\nu/q^2$. $q$ and
$M_R$ are the momentum and the mass of the fermion particle,
respectively.

\section{The production of $\Lambda_c(2940)^+$ in the proton and
antiproton collision}
\label{2940}

In this section we calculate the $\Lambda_c(2940)$ production rate in the
proton-antiproton collision as shown in Fig.~\ref{diagram22}.  For the
$p\bar{p}\to\bar{\Lambda}_c\Lambda_c(2940)^+$ process, the production
amplitudes is
\begin{eqnarray}
{\cal M}&=&g_{\Lambda_c pD^0}g_{\Lambda_c(2940)^+pD^0}
~\bar{u}_R(k_2)\mathcal{C}_R(k)u_p(p_2)\nonumber\\
&&\times\bar{v}_{  {\bar\Lambda}_c}(k_1)\mathcal{C}{v}_{\bar p}
(p_1)~G_D(k^2){\cal F}^2(k^2),\label{2}
\end{eqnarray}
where $\mathcal{C}_R$ or $\mathcal{C}$ describe the Lorentz
structures of the vertex for $D^0$ interacting with
$\Lambda_c(2940)^+ p$ or $\bar{\Lambda}_c \bar{p}$. They are
derived in terms of the Lagrangians in Eqs.~(\ref{L1})-(\ref{L6}).
$k_1$, $k_2$, $p_1$, $p_2$ and $k$ are the momenta of
$\Lambda_c(2940)^+$, $\bar{\Lambda}_c$, $p$, $\bar{p}$ and the
exchanged meson $D^0$, respectively. Additionally, the monopole
form factor ${\cal F}(k^2)=(\Lambda^2-m_D^2)/(\Lambda^2-k^2)$ is
introduced. As well understood, the concerned hadrons at the
effective vertices by no means are point-particles, but have
complicated structures, thus the  form factor phenomenologically
describes the inner structure effect of interaction vertices shown
in Fig. \ref{diagram22} and moreover, it partly compensates for the
off-shell effect of the exchanged $D^0$ meson as suggested in
Ref.~\cite{Haidenbauer:2009ad}. Indeed the monopole form factor is
a phenomenological ansatz and not derivable from the field theory,
thus errors are unavoidably brought up just like any
phenomenological computation. Since the involved parameters are
fixed by fitting data, the model-dependence is greatly alleviated,
therefore, it is observed that for lower energy reactions, the
scenario works well.

\begin{figure}[h!]
\includegraphics[bb=50 50 550 750,scale=0.36,angle=270]{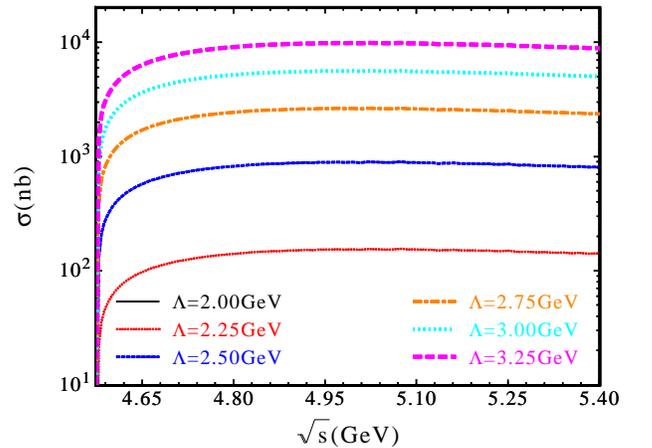}
\caption{The total cross section for the process $p\bar{p}\to
\Lambda_c\bar{\Lambda}_c$ with different $\Lambda$ values.\label{LcLc}}
\end{figure}

Before studying the cross section for the $\Lambda_c(2940)^+$
production at the $p\bar p$ collision, let us first calculate the total cross section for the
proton-antiproton scattering to the $\Lambda_c$ and anti-$\Lambda_c$ pair
in our theoretical frame, which has been experimentally measured and carefully studied in the
literature~\cite{Haidenbauer:2009ad,Goritschnig:2009sq}. In Fig.~\ref{LcLc}, the total
cross section of $p\bar{p}\to \Lambda_c\bar{\Lambda}_c$ with
different cutoffs is presented, where we restrict the $\Lambda$ value within a reasonable range from 2 GeV to 3.25 GeV.

In Ref.~\cite{Haidenbauer:2009ad}, the reaction $p\bar{p}\to\Lambda_c\bar{\Lambda}_c$
was supposed to occur via a meson-exchange mechanism, where the cutoff $\Lambda$ was set as 3 GeV. An obvious similarity between
$p\bar{p}\to\Lambda_c\bar{\Lambda}_c$ and $p\bar{p}\to\bar{\Lambda}_c\Lambda_c(2940)^+$ suggests that we adopt
$\Lambda=3$ GeV to estimate the cross section of $p\bar{p}\to\bar{\Lambda}_c\Lambda_c(2940)^+$.

\begin{figure}[h!]
\includegraphics[bb=50 50 550 750,scale=0.36,angle=270]{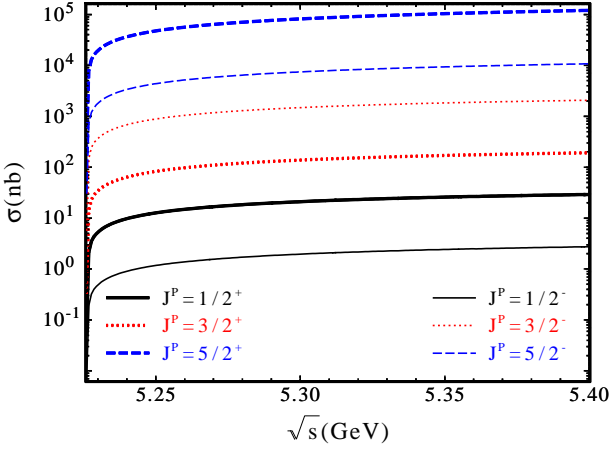}
\caption{The cross section for the process
 $p\bar{p}\to\bar{\Lambda}_c \Lambda_c(2940)^+ $ with different $J^P$ assignments of $\Lambda_c(2940)^+$.\label{Lc2940}}
\end{figure}

The cross sections for $\Lambda_c(2940)^+$ production with
different spin-parity assignments to $\Lambda_c(2940)^+$ are presented in Fig.~\ref{Lc2940}.

Our results of $\Lambda_c(2940)^+$ production indicate that the
cross section of $p\bar{p}\to \bar{\Lambda}_c\Lambda_c(2940)^+ $
strongly depends on the $J^P$ assignments of $\Lambda_c(2940)^+$.
If $\Lambda_c(2940)^+$ is  a  $J^P=1/2^-$ state, the cross section
of the $p\bar p\to \bar\Lambda_c\Lambda_c(2940)^+$ process is much
smaller than that if $\Lambda_c(2940)^+$ is a $J^P=5/2^+$ state by
a big fraction of $\sim10^4$.

\section{The Dalitz plot and the background analysis}
\label{background}

\begin{figure}[htb]
\centering
\begin{tabular}{cc}
\scalebox{0.59}{\includegraphics{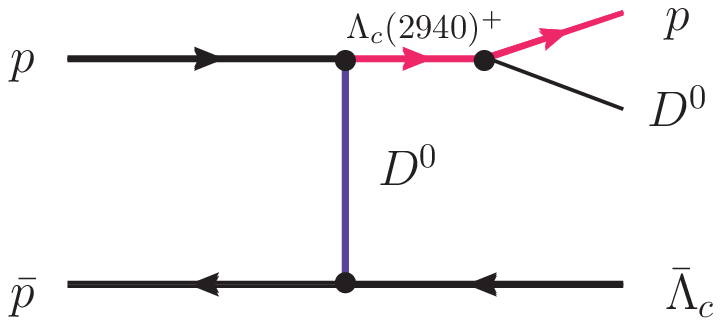}}&\scalebox{0.59}{\includegraphics{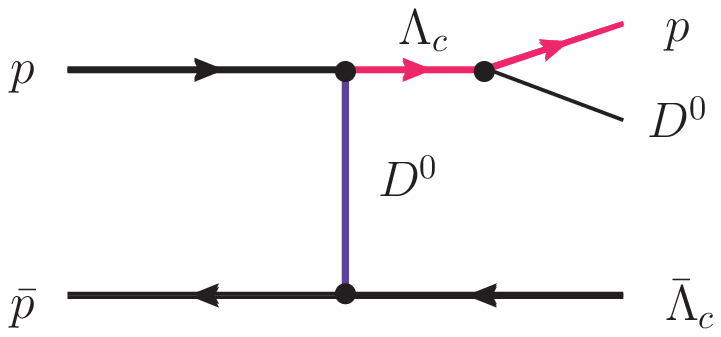}}
\end{tabular}
\caption{The diagrams for the
$p\bar{p}\to \bar{\Lambda}_c D^0 p$; the left and right diagrams occur via the intermediate $\Lambda_c(2940)^+$ and  $\Lambda_c^+$, respectively.\label{diagram}}
\end{figure}

As shown in the above section, considerable events of
$\Lambda_c(2940)^+$ can be produced in the proton and antiproton
collision. In this section, we present the  Dalitz plot of
$p\bar{p}\to \bar{\Lambda}_c D^0 p$, where $\Lambda_c(2940)$ or
$\Lambda_c$ is an intermediate state just shown in Fig.
\ref{diagram}. A comparison of Fig. \ref{LcLc} with
Fig.\ref{Lc2940} indicates that the cross section of $p\bar{p}\to
\Lambda_c\bar{\Lambda}_c$ is comparable with that  of $p\bar{p}\to
\bar{\Lambda}_c\Lambda_c(2940)^+$. Thus, $p\bar{p}\to
\Lambda_c\bar{\Lambda}_c\to \bar{\Lambda}_c D^0p$ where $\Lambda_c$ is off-shell, becomes a main
background contribution when we analyze the $\Lambda_c(2940)^+$
production in the $p\bar{p}\to \bar{\Lambda}_c\Lambda_c(2940)^+\to
\bar{\Lambda}_c D^0 p$ process.

The amplitude of $p\bar{p}\to \bar{\Lambda}_c\Lambda_c(2940)^+\to \bar{\Lambda}_c D^0 p$ where $\Lambda_c(2940)$ can be an on-shell baryon, reads as
\begin{eqnarray}
    {\cal M}&=&g_{\Lambda_cpD^0}g_{\Lambda_c(2940)^+pD^0}^2\bar{u}_p(k_2)\Gamma_R(k_3)G^{n+\frac{1}{2}}_R(q)
    \Gamma_R(k)u_p(p_2)\nonumber\\
    &&\times\bar{v}_{{\bar\Lambda}_c}(k_1)\Gamma{v}_{\bar p} (p_1)
    G_D(k^2){\cal F}^2(k^2),\label{ha}
\end{eqnarray}
where $q$, $k_2$ and $k_3$ are the four-momenta of the
intermediate state $\Lambda_c(2940)^+$ and final states $p$ and
$D^0$, respectively. We can easily obtain the amplitude of
$p\bar{p}\to \Lambda_c\bar{\Lambda}_c\to \bar{\Lambda}_c D^0p$ by
Eq. (\ref{ha}), where we only need to replace the relevant
parameter of $\Lambda_c(2940)^+(J^P=1/2^+)$ with that of
$\Lambda_c$.

\begin{figure}[h!]
\includegraphics[bb=50 50 550 750,scale=0.36,angle=270]{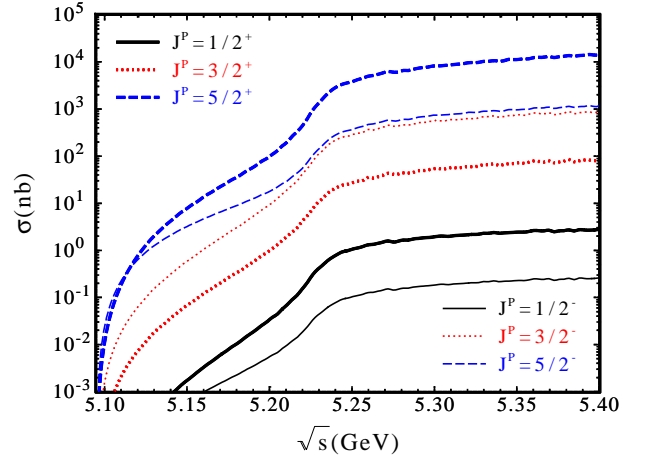}
\caption{The dependence of the cross section for the $p\bar{p}\to \bar{\Lambda}_c\Lambda_c(2940)^+\to \bar{\Lambda}_c D^0 p$ process
on $\sqrt{s}$. Here, we consider different $J^P$ assignments to
$\Lambda_c(2940)^+$.\label{Lc}}
\end{figure}

In Fig.~\ref{Lc}, we present the cross section of $p\bar{p}\to
\bar{\Lambda}_c\Lambda_c(2940)^+\to \bar{\Lambda}_c D^0 p$, which
is dependent on $\sqrt{s}$. As shown in Fig. \ref{Lc}, there
exists a steep increase at about $\sqrt{s}=5.2$ GeV, where
$\Lambda(2940)^+$ approaches its mass-shell, so its propagator
contributes a cusp.

\begin{figure}[h!]
\includegraphics[bb=50 50 550 750,scale=0.36,angle=270]{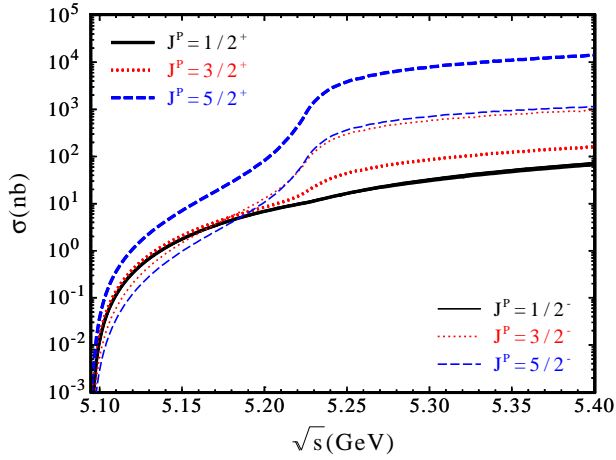}
\caption{The cross section of $p\bar{p}\to \bar{\Lambda}_c D^0 p$. Here, we include the background contribution to $p\bar{p}\to \bar{\Lambda}_c D^0 p$. \label{Lct}}
\end{figure}

Taking the background contribution into account, the dependence of
the cross section of $p\bar{p}\to \bar{\Lambda}_c D^0 p$ on $\sqrt
s$ is shown in Fig.~\ref{Lct}. Our calculation also indicates that
the order of magnitude of the cross section of $p\bar{p}\to
\Lambda_c\bar{\Lambda}_c\to \bar{\Lambda}_c D^0p$ is about 10 nb,
which is far larger than that of
 $p\bar{p}\to \Lambda_c(2940)^+\bar{\Lambda}_c\to \bar{\Lambda}_c D^0p$ as $\Lambda_c(2940)^+$ is a $J^P=1/2^+$
state. To some extent,  the
contribution of the intermediate $\Lambda_c(2940)^+$ of $J^P=1/2^-$ to $p\bar{p}\to  \bar{\Lambda}_c D^0p$ is immersed in the background.

\begin{figure}[h!]
\includegraphics[bb=40 310 520 720,scale=0.25,clip]{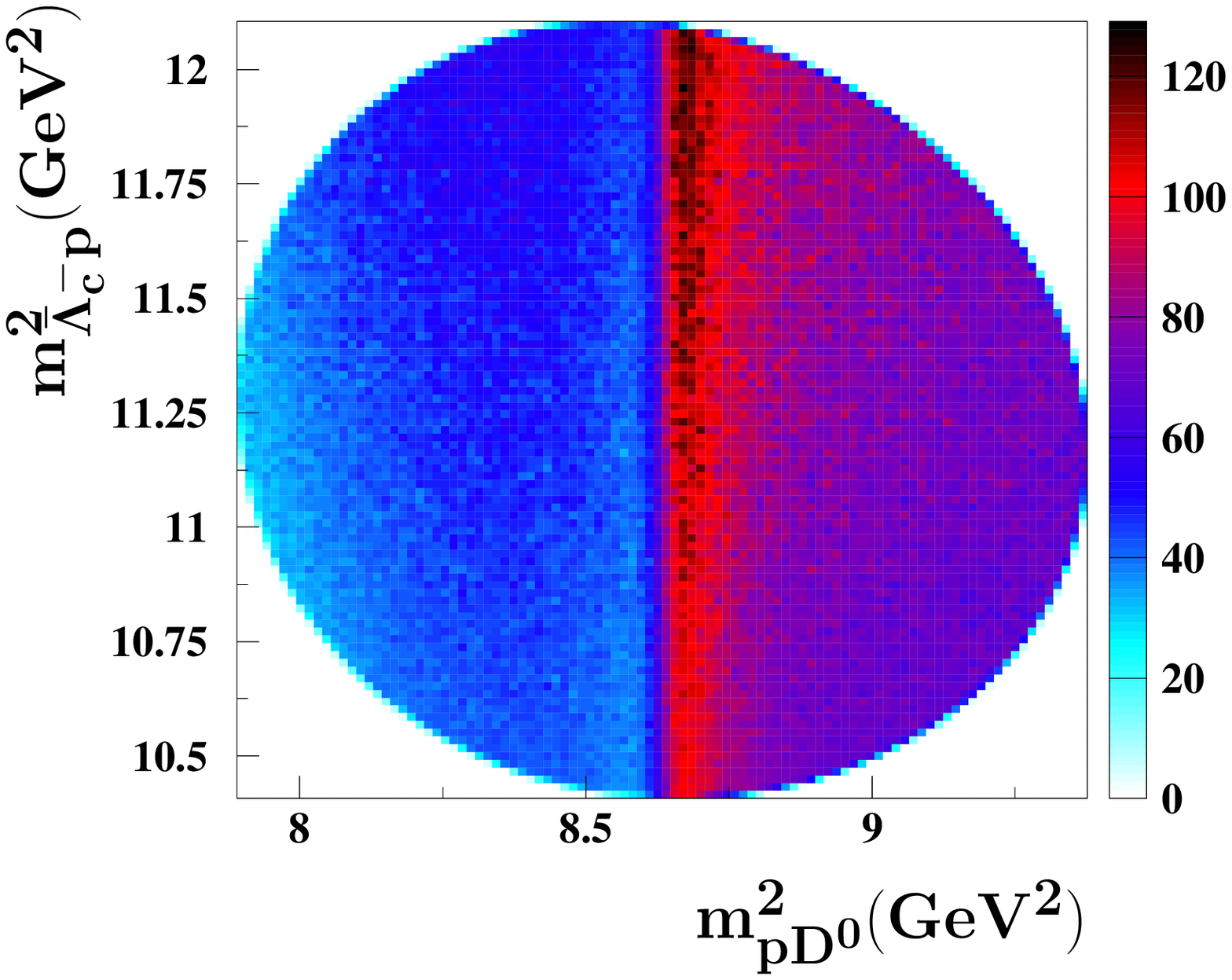}
\includegraphics[bb=40 310 520 720,scale=0.25,clip]{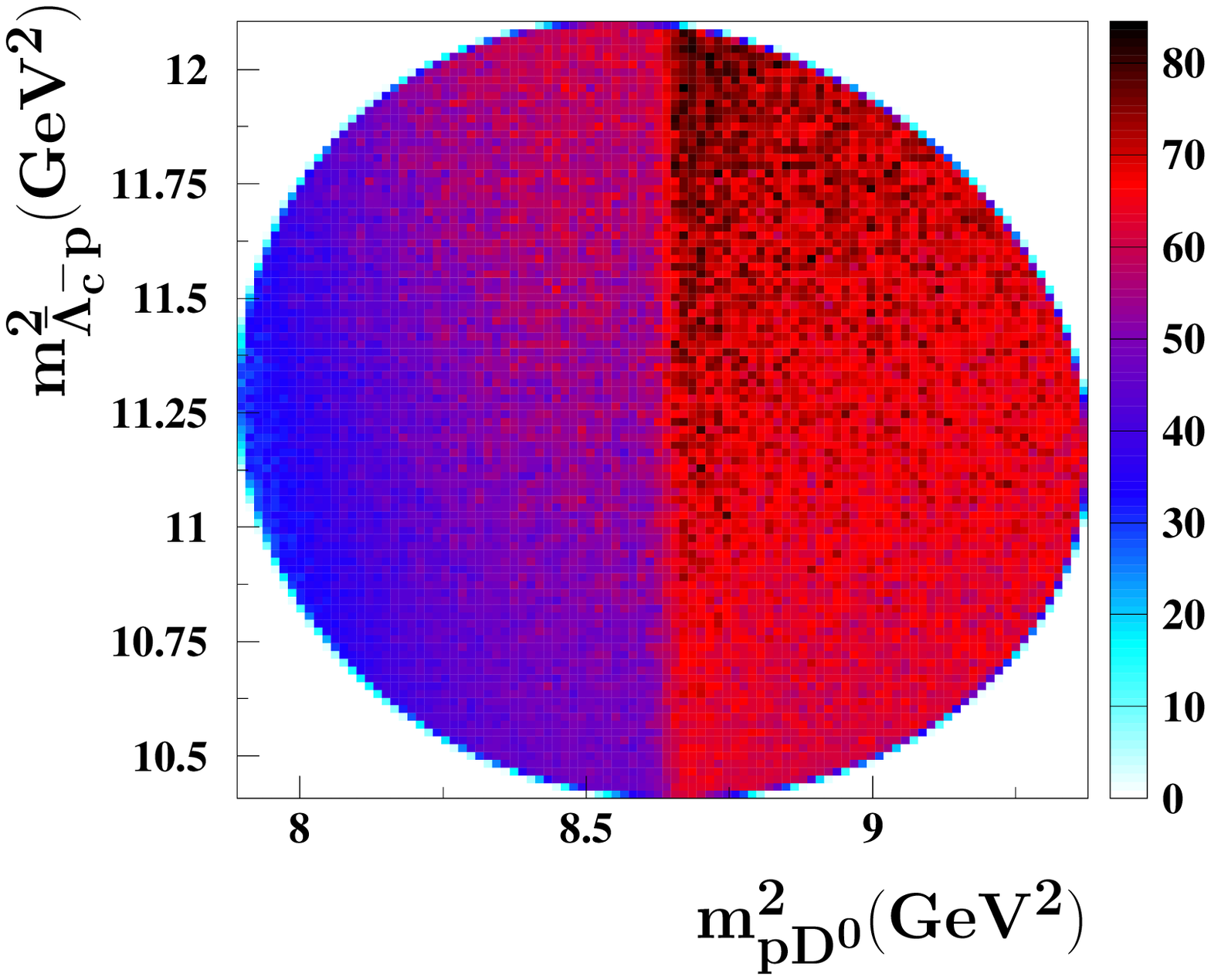}\\
\includegraphics[bb=40 310 520 720,scale=0.25,clip]{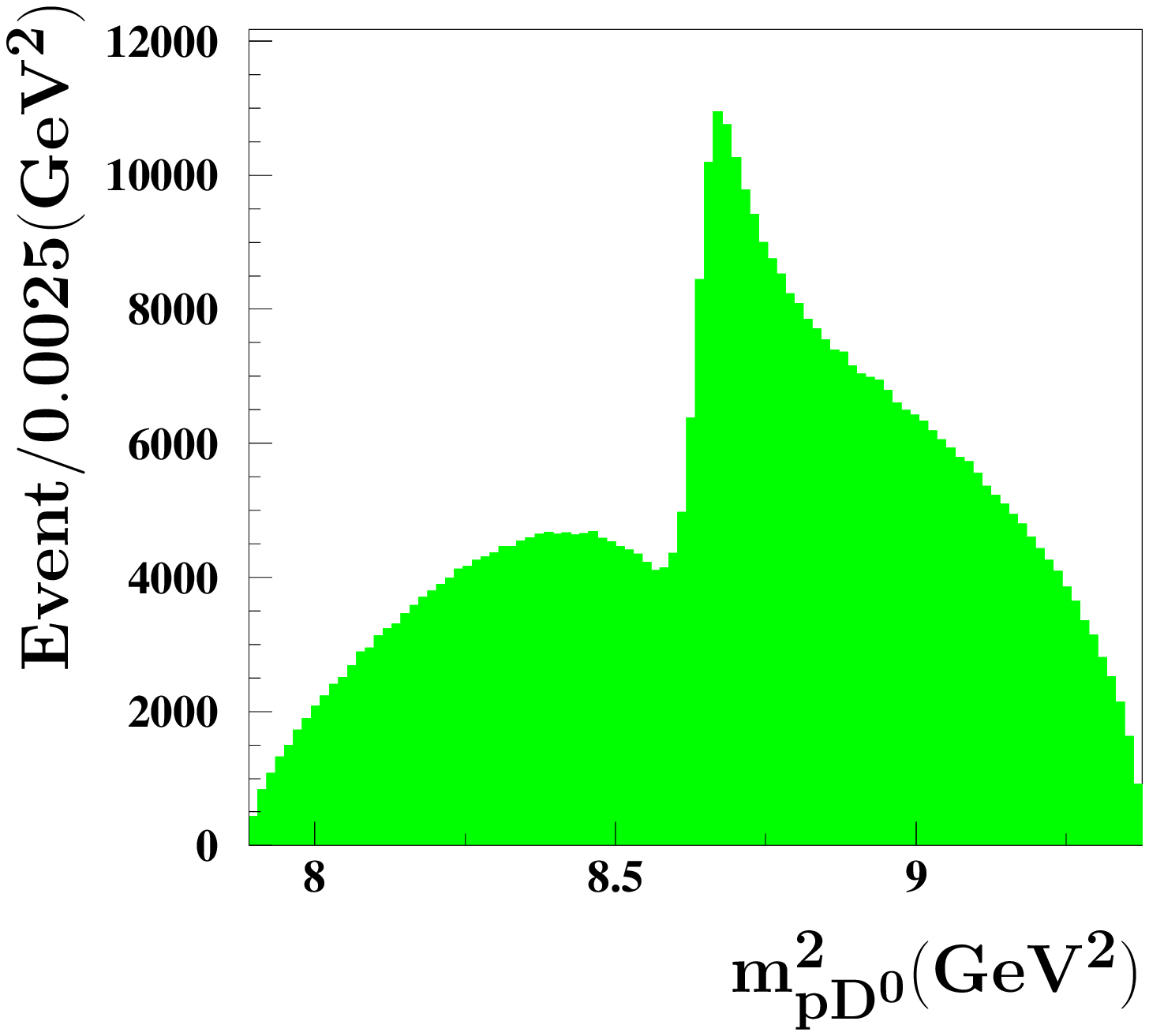}
\includegraphics[bb=40 310 520 720,scale=0.25,clip]{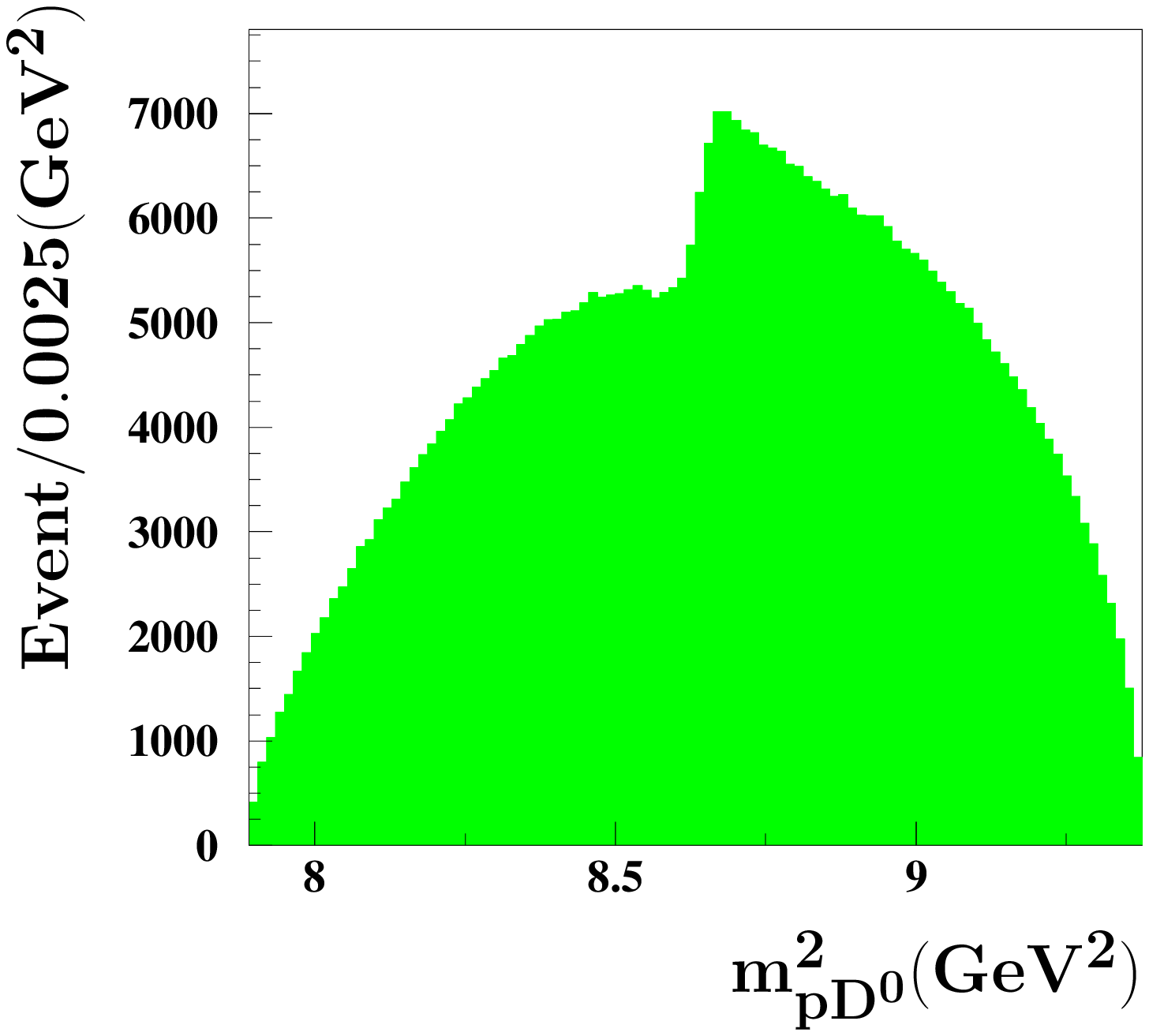}
\caption{The Dalitz plot and invariant mass spectra for $p\bar{p}\to \bar{\Lambda}_c D^0 p$
at $\sqrt{s}=5.32$~GeV and with $J=1/2$ assignment to $\Lambda_c(2940)^+$. Here, the left or right
column corresponds to the numerical result of the production of $\Lambda_c(2940)^+$ with positive or negative parity.\label{Dalitz1}}
\end{figure}

\begin{figure}[h!]
\includegraphics[bb=40 310 520 720,scale=0.25,clip]{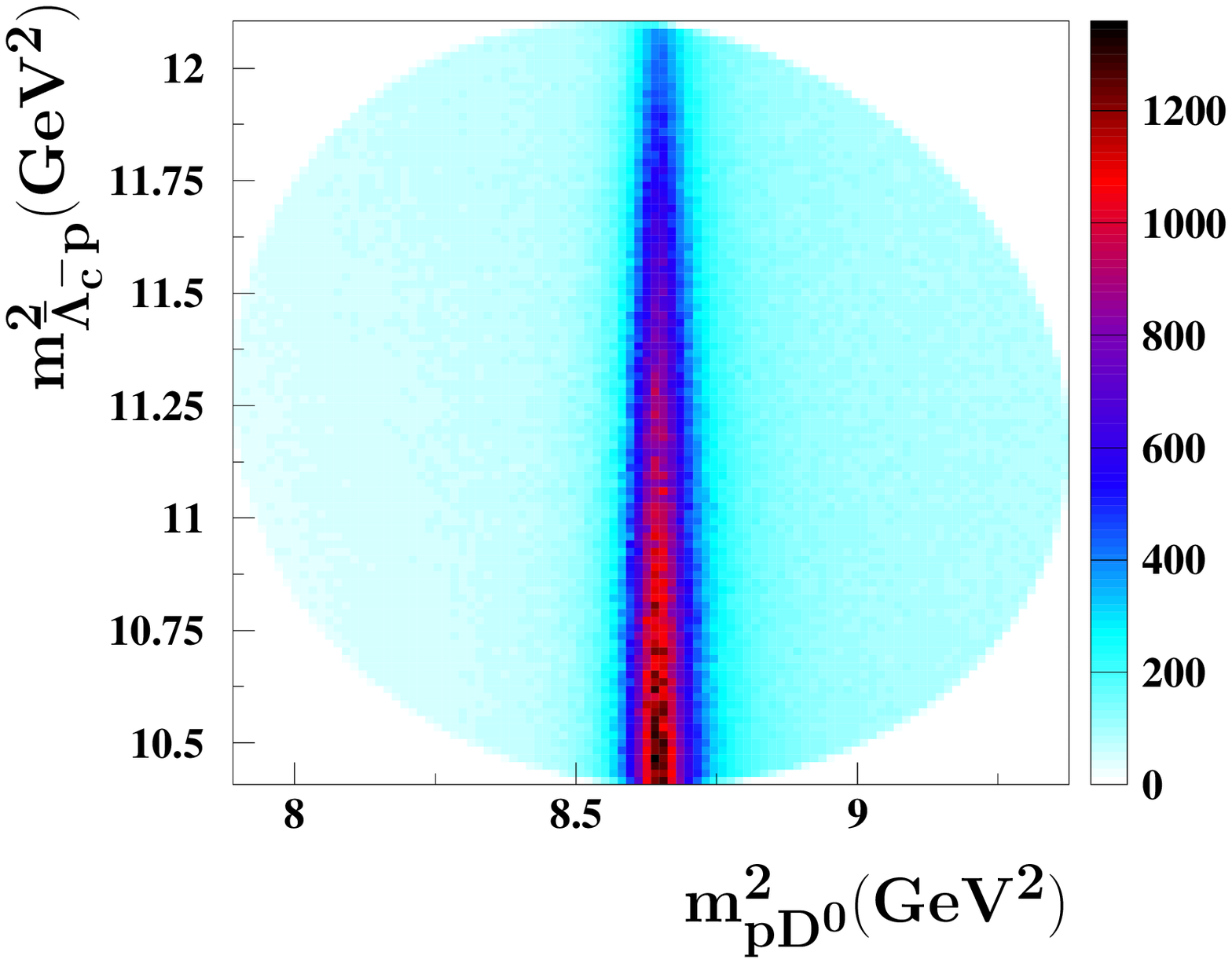}
\includegraphics[bb=40 310 520 720,scale=0.25,clip]{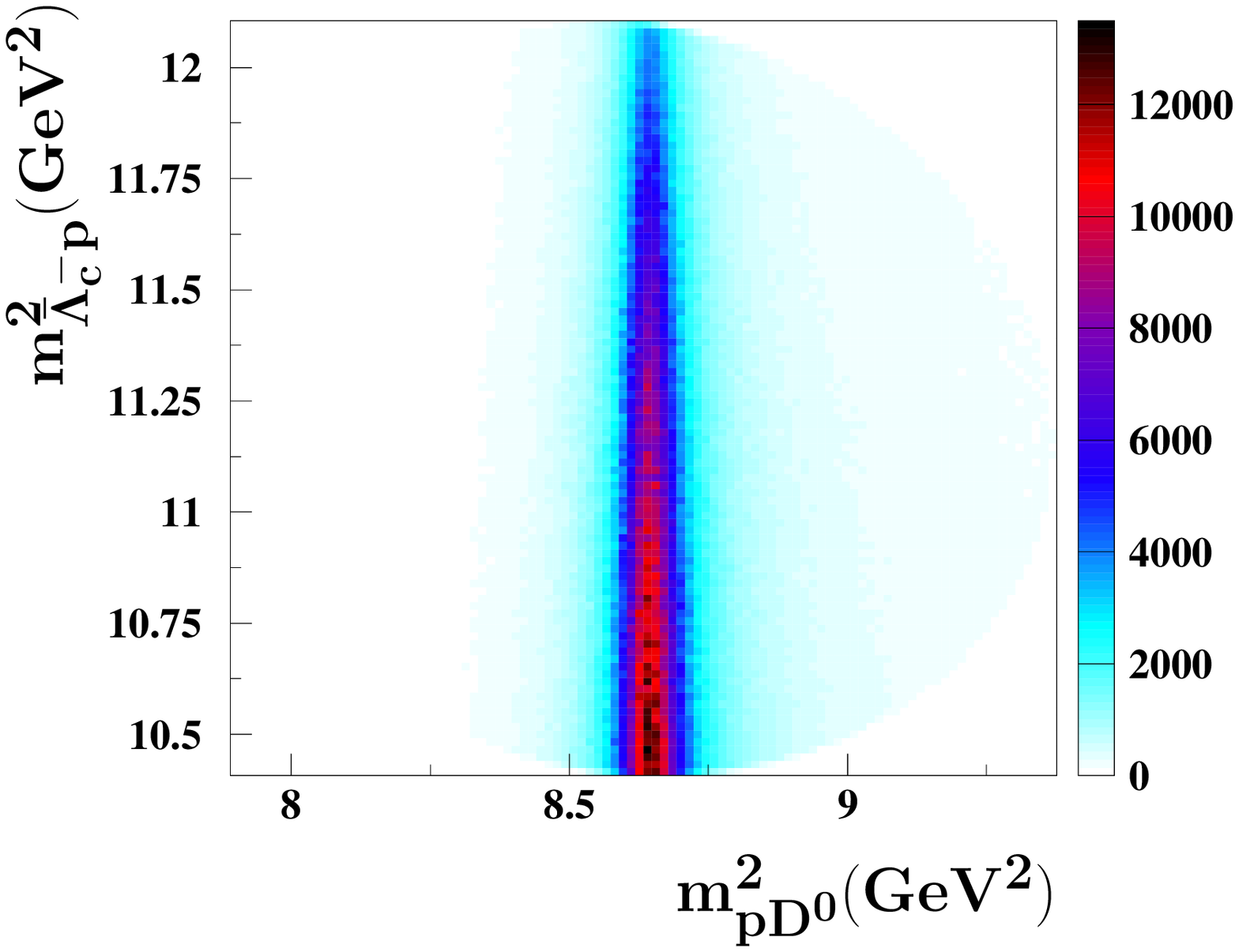}\\
\includegraphics[bb=40 310 520 720,scale=0.25,clip]{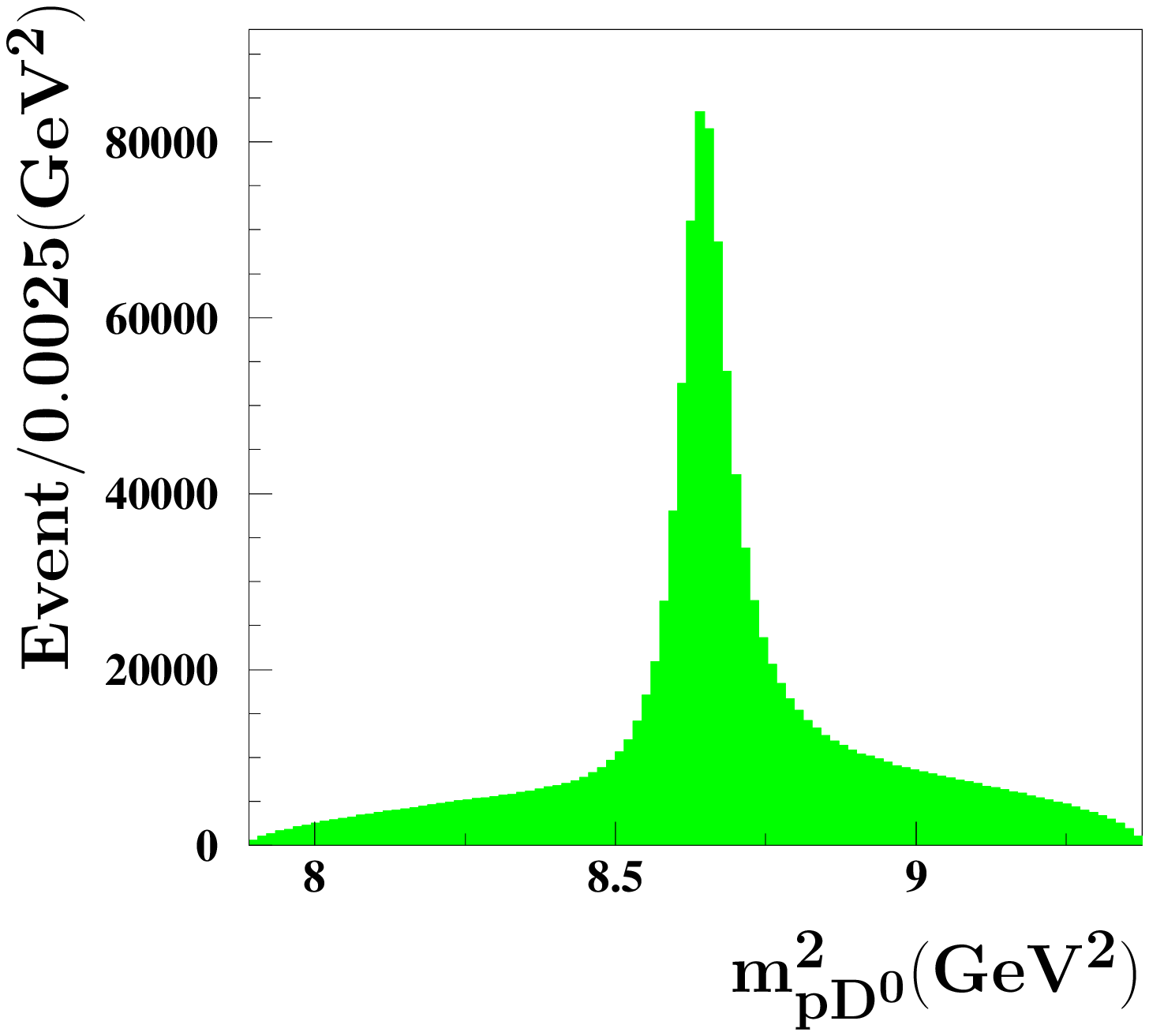}
\includegraphics[bb=40 310 520 720,scale=0.25,clip]{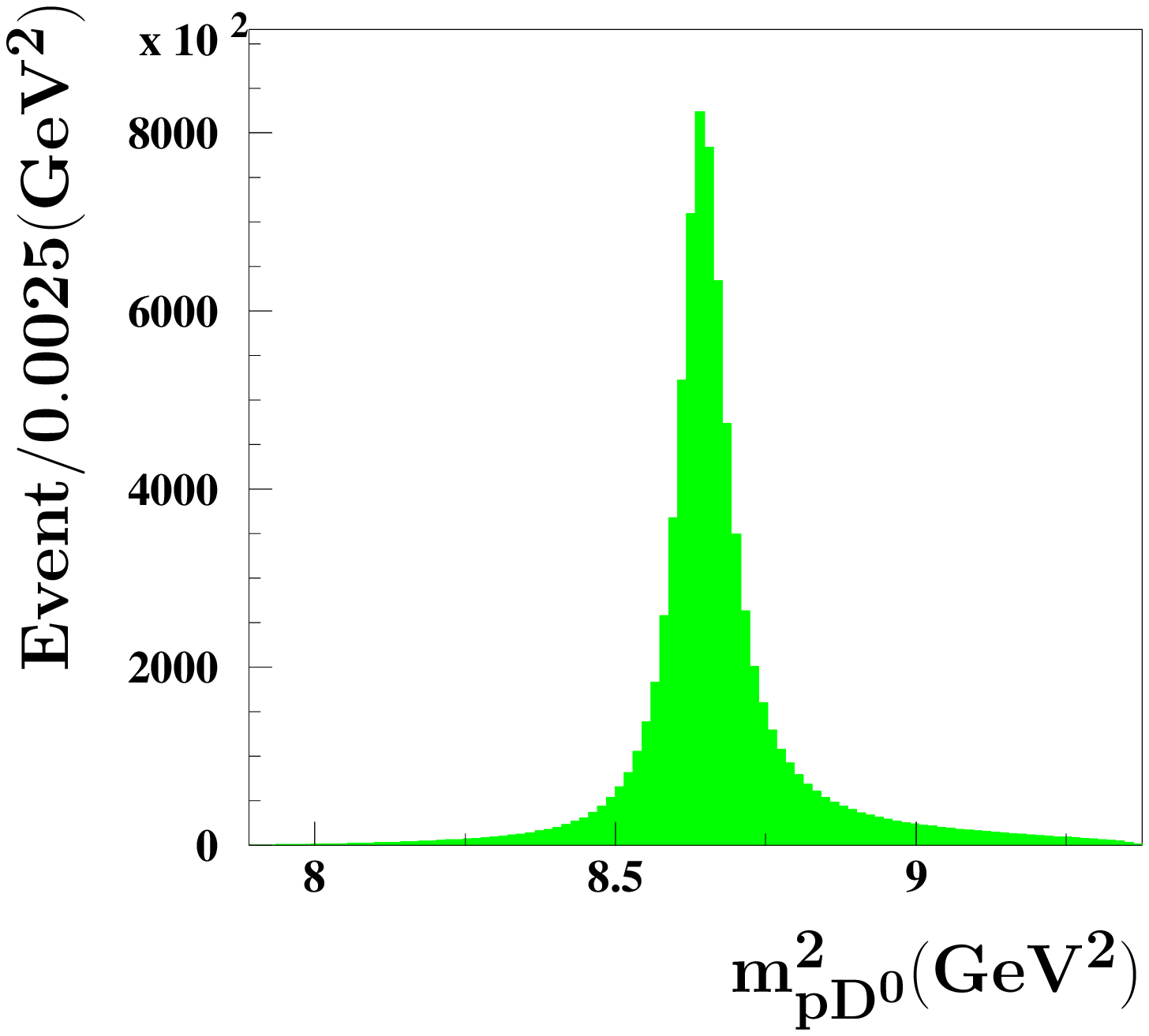}
\caption{The Dalitz plot and invariant mass spectra for $p\bar{p}\to \bar{\Lambda}_c D^0 p$
at $\sqrt{s}=5.32$~GeV and with $J=3/2$ assignment to $\Lambda_c(2940)^+$. Here, the left or right
column corresponds to the numerical result of the production of $\Lambda_c(2940)^+$ with positive or negative parity.\label{Dalitz2}}
\end{figure}

\begin{figure}[h!]
\includegraphics[bb=40 310 520 720,scale=0.25,clip]{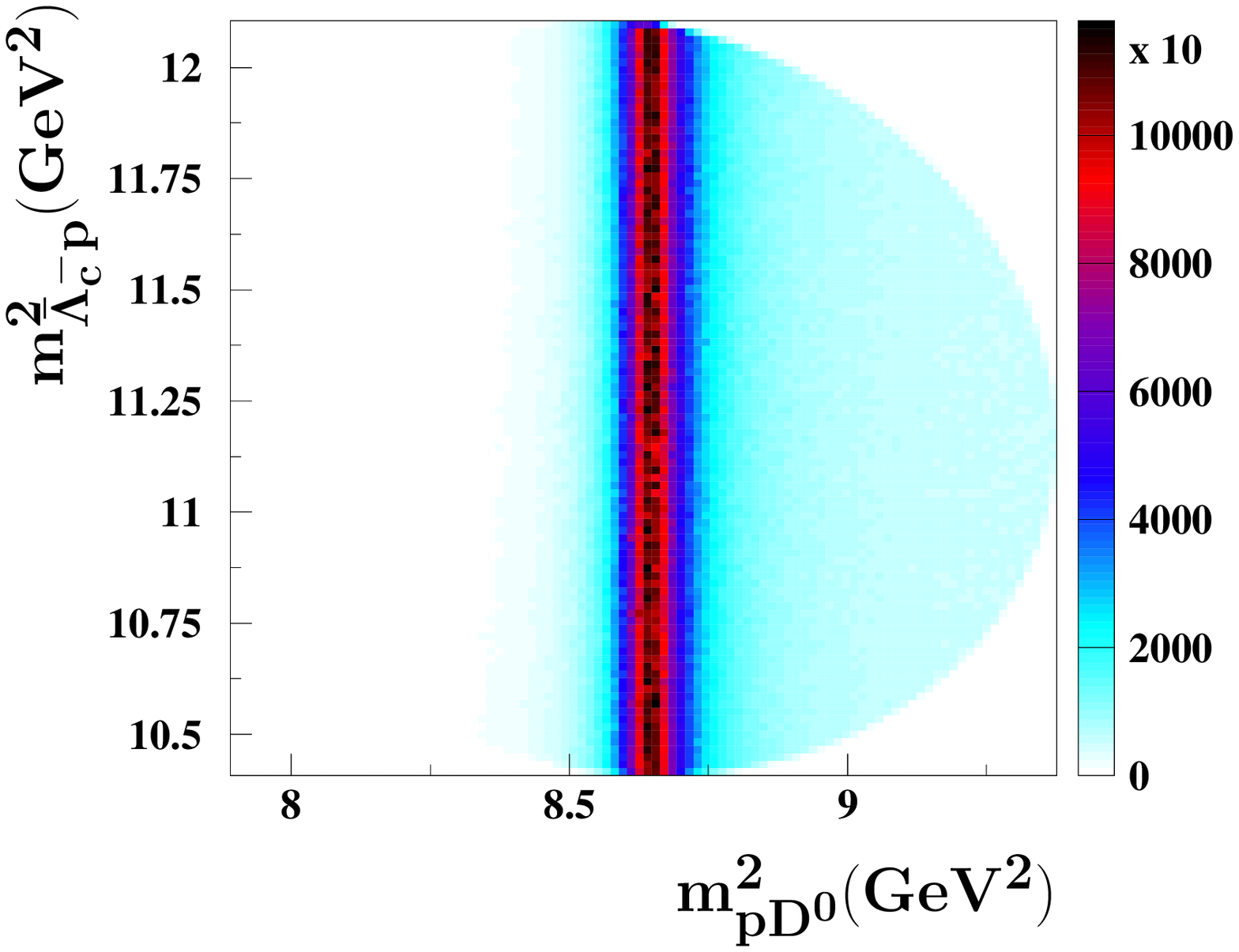}
\includegraphics[bb=40 310 520 720,scale=0.25,clip]{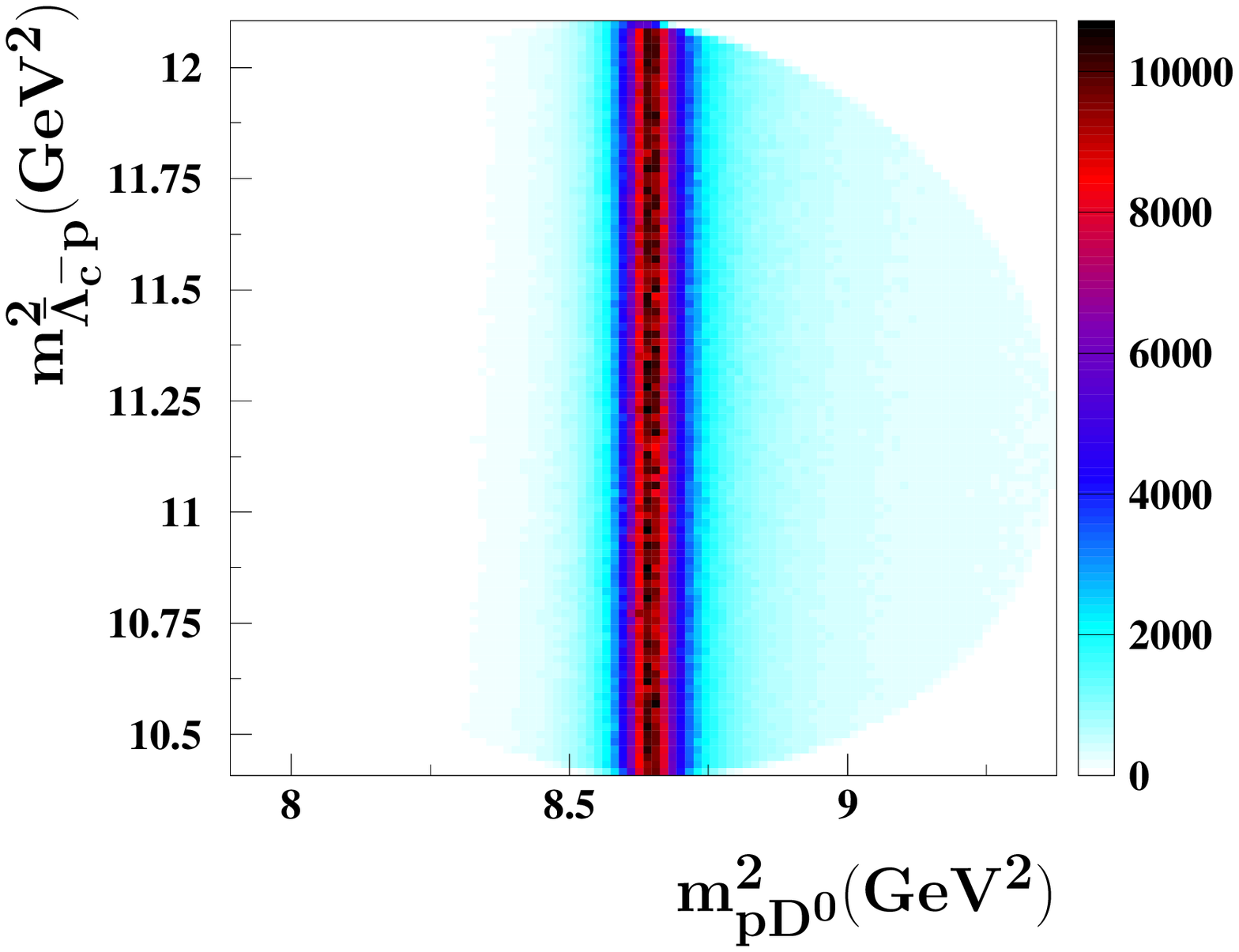}\\
\includegraphics[bb=40 310 520 720,scale=0.25,clip]{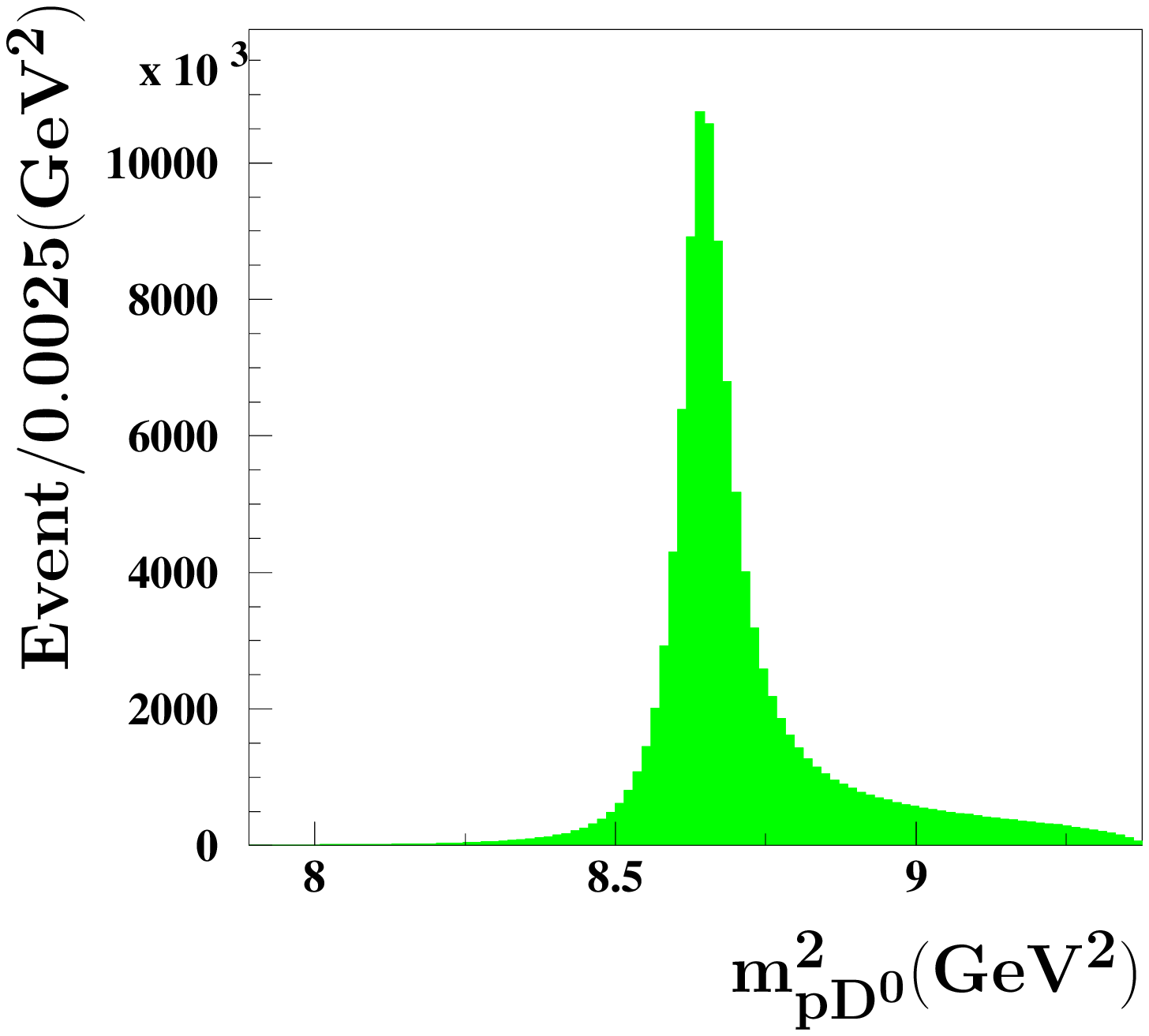}
\includegraphics[bb=40 310 520 720,scale=0.25,clip]{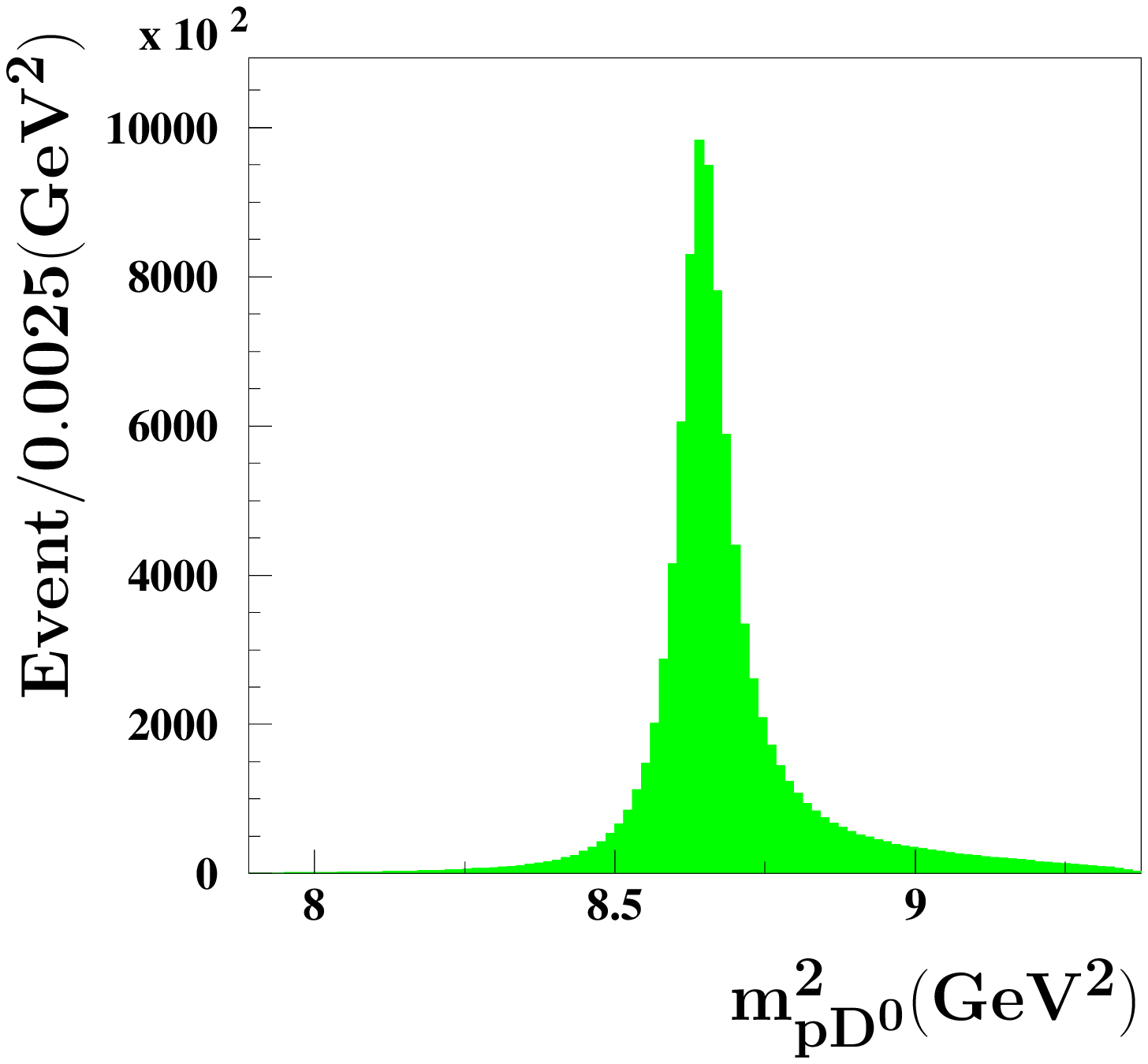}
\caption{The Dalitz plot and invariant mass spectra for $p\bar{p}\to \bar{\Lambda}_c D^0 p$
at $\sqrt{s}=5.32$~GeV and with $J=5/2$ assignment to $\Lambda_c(2940)^+$. Here, the left or right
column corresponds to the numerical result of the production of $\Lambda_c(2940)^+$ with positive or negative parity.\label{Dalitz3}}
\end{figure}

The Dalitz plot is a very useful tool for the data analysis since much
information is exposed by the plot.
With the help of the FOWL code, we present the Dalitz plot for the
$p\bar{p}\to \bar{\Lambda}_c D^0 p$ process and the $pD^0$ invariant mass spectrum $m^2_{pD^0}$ in Figs. \ref{Dalitz1}-\ref{Dalitz3}.

Just as shown in Fig. \ref{Dalitz1}, the shape of the
distributions, where peaks appear at certain locations, are not
the Breit-Wigner types. This is mainly due to an interference
between the amplitudes of $p\bar{p}\to \Lambda_c\bar{\Lambda}_c\to
\bar{\Lambda}_c D^0p$ and $p\bar{p}\to
\Lambda_c(2940)^+\bar{\Lambda}_c\to \bar{\Lambda}_c D^0p$, which
also implies that  $p\bar{p}\to \Lambda_c\bar{\Lambda}_c\to
\bar{\Lambda}_c D^0p$ forms the dominant background for
$p\bar{p}\to \bar{\Lambda}_c D^0p$.


With $J^P=3/2^\pm$ or $5/2^\pm$ assignments to
$\Lambda_c(2940)^+$, we find that there exist explicit cusp
structures corresponding to $\Lambda_c(2940)^+$ in the $pD^0$
invariant mass spectrum, which can be described by the
Breit-Wigner formalism. The Dalitz plot analysis indicates that
$\Lambda_c(2940)^+$ signal can be well distinguished from the
background in the $p\bar{p}\to \bar{\Lambda}_c D^0 p$ process.
That is due to the fact that the contribution of $p\bar{p}\to
\Lambda_c\bar{\Lambda}_c\to \bar{\Lambda}_c D^0p$ is far smaller
than that of $p\bar{p}\to \Lambda_c(2940)^+\bar{\Lambda}_c\to
\bar{\Lambda}_c D^0p$ as shown in Figs. \ref{Lc}-\ref{Lct}.

\section{Discussion and conclusion}\label{dis}

In this work we investigate the production rate of
$\Lambda_c(2940)^+$ in the future experiments at PANDA. We find if the
branching ratio of $\Lambda_c(2940)^+$ decaying into $D^0p$
is at the order 0.1, at least $10^4$ events of
$\Lambda_c(2940)^+$ per day can be produced at PANDA.

Here, let us briefly discuss dependence of the numerical result on the phenomenologically introduced
parameter $\Lambda$ used in this work. The
cutoff $\Lambda=3$ GeV is adopted as suggested in
Ref.~\cite{Haidenbauer:2009ad}. If the
cutoff $\Lambda$ decreases to $2.5$ GeV, both the production rate of $\Lambda_c(2940)^+$ and the
background would increase about one order. The number of events
is still large enough for investigating behaviors of $\Lambda_c(2940)^+$
in the proton and antiproton collision. In our numerical computations we adopt the
same cutoff $\Lambda$ value as that in Ref. \cite{Haidenbauer:2009ad}.

\begin{figure}[htb]
\centering
\begin{tabular}{c}
\scalebox{1.3}{\includegraphics{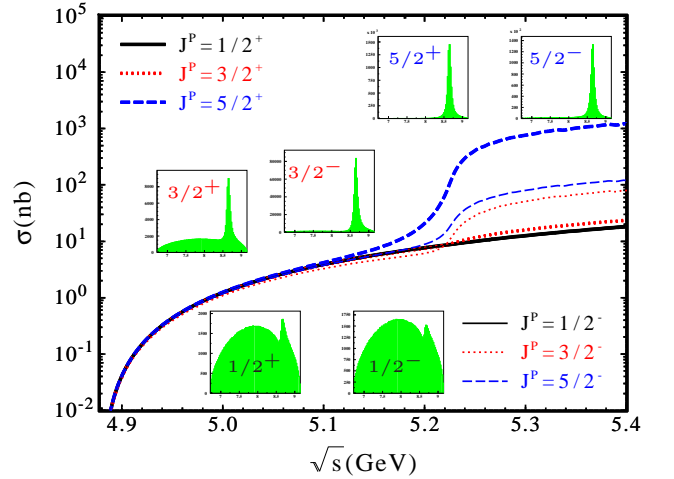}}
\end{tabular}
\caption{The total cross section and invariant mass spectrum for $p\bar{p}\to
\pi^-\Sigma_c^{++}\bar{\Lambda}^-_c$
at $\sqrt{s}=5.35$~GeV. Here, we consider the ISI effect just discussed in Sec. \ref{dis}. \label{Sp}}
\end{figure}

We would like to specify an important issue, which was discussed in literature and may affect our
theoretical estimate of the production rate. It is noted that the initial state interaction (ISI)
effect is included in the numerical result presented in Secs. \ref{2940} and \ref{background}.
The ISI is an important effect for studying meson production in nucleon-nucleon
collisions as the transition occurs near the threshold. That effect was first observed by the authors of Refs. \cite{Baru:2002rs,Hanhart:2003pg} that
the ISI makes the cross section to be reduced by an overall factor, which is slightly energy-dependent.
In studying $p\bar{p}\to\bar{\Lambda}_c\Lambda_c$ process, the authors of Ref. \cite{Haidenbauer:2009ad} also took into account  the ISI effect, which reduces
the cross section of $p\bar{p}\to\bar{\Lambda}_c\Lambda_c$ by a factor 10. The ISI may be induced by complicated interaction processes among the ingredients
inside the colliding $p$ and $\bar p$, which may be valence quarks or even gluons and sea quarks. It is believed that such processes  are governed by
the non-perturbative QCD effects, thus not calculable so far. Interesting to note that for  high energy $p\bar p$ or $pp$ collisions, one can use the parton
distribution function (PDF) due to the asymptotic freedom of QCD, but for lower energy collisions, we do not know how to correctly deal with the ISI effects.
Therefore, as suggested by previous research \cite{Baru:2002rs,Hanhart:2003pg}, we would retain a phenomenological factor in the numerical estimate of the production rate to take care of
the ISI effect on $p\bar{p}\to\bar{\Lambda}_c\Lambda_c(2940)^+$. Thus, an extra factor is introduced to reflect the ISI effect,
which makes the cross section of $p\bar{p}\to\bar{\Lambda}_c\Lambda_c(2940)^+$ corresponding to Eq. (\ref{2})
suppressed by one order of magnitude (the ISI effect is considered in the numerical results presented in Figs. \ref{LcLc}-\ref{Sp}).
With above consideration, we can roughly estimate the production events
of $\Lambda_c(2940)^+$ at PANDA and the results are presented in Fig. \ref{Lc2940}. The designed luminosity of PANDA is about
$2\times10^{32}$ cm$^{-2}$/s, so the integrated luminosity in one day run
is about $10^4$ nb$^{-1}$. Assuming we have a 50\% overall efficiency, we may expect
$10^4\sim10^8$ events of $\Lambda_c(2940)^+$ per day produced at PANDA. In addition, we also would like to emphasize that the qualitative
conclusion, which is made via the background analysis and Dalitz plot, is not affected by
whether including the ISI effect.

Furthermore, the line shape of the cross
section and invariant mass spectrum without taking in the background is independent of the coupling
constant  $g_{\Lambda_c(2940)^+D^0p}$.
If the branching ratio of $\Lambda_c(2940)^+\to D^0p$ is about 10\%, there is
a large final-state phase space for the production of $p\bar{p}\to \Lambda_c\bar{\Lambda}_c\to \bar{\Lambda}_c D^0p$. As
$10^4\sim10^8$ of $\Lambda_c(2940)^+$ per day can be produced, one can carefully study the properties of
$\Lambda_c(2940)^+$ via the channel of $p\bar p\to \Lambda_c(2940)+\bar\Lambda_c\to D^0p+\bar D^0\bar p$ in the future PANDA experiments.
In the second sub-process, $\Lambda_c$ decays into $\bar D^0+\bar p$
which is easy to be experimentally observed and the constructed invariant mass can accurately identify $\Lambda_c$.

Since the Belle Collaboration confirmed $\Lambda_c(2940)^+$ in
the $\Sigma_c(2455)^{0,++}\pi^{+,-}$ channels \cite{Abe:2006rz}, we also study the $\Lambda_c(2940)^+$ production in
$p\bar{p}\to \pi^-\Sigma^{++}\bar{\Lambda}^-_c$, where $p\bar{p}\to \bar{\Lambda}_c\Lambda_c\to \bar{\Lambda}^-_c \pi^-\Sigma_c^{++}$
and $p\bar{p}\to \bar{\Lambda}_c\Lambda_c(2940)^+\to \bar{\Lambda}_c \pi^-\Sigma_c^{++}$
compose the background and signal for the $\Lambda_c(2940)^+$ production respectively. In the former channel, because of constraint from the phase space, the $\Lambda_c$ can only be an off-shell intermediate state for the final state  $\pi\Sigma_c$, so due to the Breit-Wigner structure, such sub-process is relatively suppressed in comparison with the "signal".
The cross section and the invariant mass spectrum of $p\bar{p}\to
\bar{\Lambda}_c\pi^-\Sigma^{++}$
with $\sqrt{s}=5.35$~GeV and
$B(\Lambda_c(2940)^+\to\pi^-\Sigma_c^{++})\sim$10\% is presented in Fig.~\ref{Sp}. Here, we take the
coupling constant as $g_{\Lambda_c\Sigma_c\pi}=3.9$, which results in a weaker
background. The signals of $\Lambda_c(2940)^+$ can be distinguished from the
background easily as shown in Fig.~\ref{Sp}. Thus, one can conclude that the channel $p\bar{p}\to
\pi^-\Sigma_c^{++}\bar{\Lambda}_c$ is also a suitable channel to study $\Lambda_c(2940)^+$.

Based on the analysis above, it is optimistic to investigate
$\Lambda_c(2940)^+$ in the future experiment of PANDA, even though the cross section is not as large as for
the charminium-like states, such as $X(3872)$ \cite{Chen:2008cg}.

In addition, it is very interesting to notice the observation potential at BelleII \cite{belleii,Aushev:2010bq} and the Super$B$ factory \cite{superb}, which will produce
a large database of $\Upsilon(4S)$. As $\Upsilon(4S)$ may have a sizable branching ratio to decay into $\Lambda_c(2940)+\bar \Lambda_c(2940)
(\bar \Lambda_c)$, thus comparing the data obtained at PANDA with that from the $B$-factory would make more sense and help eventually to pin down the spin-parity assignment of $\Lambda_c(2940)^+$.

\section*{Acknowledgement}

This project is supported by the National Natural Science
Foundation of China under Grants No. 11175073, No. 10905077, No.
11005129, No. 11035006, No. 11075079; the Ministry of Education of
China (FANEDD under Grant No. 200924, DPFIHE under Grant No.
20090211120029, NCET under Grant No. NCET-10-0442); the project
sponsored by SRF for ROCS, SEM under Grant No. HGJO90402 and the
Special Foundation of President Support by the Chinese Academy of
Sciences under Grant No. YZ080425.

\end{document}